# A review of multiscale 0D-1D computational modeling of coronary circulation with applications to cardiac arrhythmias

Stefania Scarsoglio[1*], Luca Ridolfi[2]

[1] Department of Mechanical and Aerospace Engineering, Politecnico di Torino, Torino, Italy

[2] Department of Environmental, Land and Infrastructure Engineering, Politecnico di Torino, Torino, Italy

*Corresponding author: Stefania Scarsoglio, stefania.scarsoglio@polito.it

Running title: 0D-1D computational models of coronary circulation

## Abstract

Computational hemodynamics is becoming an increasingly important tool in clinical applications and surgical procedures involving the cardiovascular system. Aim of this review is to provide a compact summary of state of the art 0D-1D multiscale models of the arterial coronary system, with particular attention to applications related to cardiac arrhythmias, whose effects on the coronary circulation remain so far poorly understood. The focus on 0D-1D models only is motivated by the competitive computational cost, the reliability of the outcomes for the whole cardiovascular system, and the ability to directly account for cardiac arrhythmias. The analyzed studies show that cardiac arrhythmias by their own are able to promote significant alterations of the coronary hemodynamics, with a worse scenario as the mean heart rate (HR) increases. The present review can stimulate future investigation, both in computational and clinical research, devoted to the hemodynamic effects induced by cardiac arrhythmias on the coronary circulation.

## 1. Introduction

By combining numerical techniques and mathematical modeling of different order and geometric detail, computational hemodynamics is becoming a powerful tool in translational medicine to accurately reproduce the human cardiovascular system [1-4]. The computational approach allows to analyze hemodynamic variables difficult to be measured, understand physiological and pathological mechanisms, simulate the hemodynamic response induced by medical procedures (e.g., drug administration and surgery), monitor the onset and progress of cardiovascular diseases, and obtain patient-specific hints.

Here we focus on the coronary circulation modeling in presence of cardiac arrhythmias, in particular atrial fibrillation (AF). Coronary artery disease (CAD), involving a reduction of blood flow supply to the heart, is one of the main cardiovascular disease and the major cause of death worldwide [5]. AF is the most common sustained clinical arrhythmia, inducing irregular and accelerated heart beating, with increasing incidence and prevalence especially in the elderly [6]. CAD and AF frequently coexist, even if definitive clinical data regarding the hemodynamic effects of AF on coronary circulation are still missing [7, 8]. Thus, computational hemodynamics can be a valuable tool in clinical applications.

Coronary blood flow has been recently modeled ranging from fully 3D approaches coupled with electro-mechanics response to lower order multiscale models [9-12]. Among the different approaches, aim of this review is to provide a compact summary of state of the art 0D-1D multiscale models of the arterial coronary system. The focus on 0D-1D models only is motivated by the fact that these multiscale models guarantee a very good compromise between the computational cost, the level of the hemodynamic details, and the reliability of the outcomes [13]. 0D-1D multiscale models can be easily coupled with 3D models of specific cardiovascular regions, and are able to capture wave transmission and reflection phenomena along the coronary network. In the end, 0D-1D models are well suited to study the effects of cardiac arrhythmias on the coronary circulation, which is the object of the present review.

## 2. Methods

The typical scheme adopted for 0D-1D coronary blood flow modeling is reported in Fig. 1 and here briefly described (symbols and abbreviations are defined in Table 1):

- Arterial coronary circulation. The three main coronary branches, right coronary artery (RCA), left anterior descending artery (LAD), and left circumflex artery (Cx) are usually modeled as 1D vessels [2, 13, 14]. Depending on the level of details of the specific model, further coronary arteries and related branches (as shown in Fig. 1) are as well modeled as 1D vessels. As the 1D modeling presents some fundamental features which are common to most part of the literature, we define as classical the 1D model composed by these basic elements: vessels are considered axisymmetric, whose walls are impermeable, tapered, longitudinally-tethered, and only subject to small and radial deformations. Flow is laminar and pressure is considered uniform across the section. Blood is assumed Newtonian, incompressible, and characterized by constant density and kinematic viscosity. Effects of suspended particles are neglected.

Under these hypotheses, the 1D model is composed by continuity and momentum equations in one-dimensional form, obtained as the integro-differential form of the balance equations of mass and momentum, and assuming: (i) flow properties as function of a single spatial variable (i.e., the vessel axial coordinate $x$), and (ii) uniform flow properties over the vessel cross-sectional area $A$, normal to the axial coordinate $x$. In so doing, $x$ and $t$ (time) are the independent variables, while area ($A$), pressure ($P$), and flow rate ($Q$) are the dependent variables. A constitutive equation - being linear or non-linear to account for the vessel elasticity or viscoelasticity respectively - links $P$ and $A$, along the $x$ direction and completes the differential system. The general form of the system of hyperbolic partial differential equations reads

$$(1) \quad \begin{cases} \dfrac{\partial A}{\partial t} + \dfrac{\partial Q}{\partial x} = 0 \\ \dfrac{\partial Q}{\partial t} + \dfrac{\partial}{\partial x}\left(\dfrac{Q^2}{A}\right) = -\dfrac{A}{\rho}\dfrac{\partial P}{\partial x} - 2\pi r_i \dfrac{\tau}{\rho} \\ P = f(A) \end{cases}$$

where $\tau$ is the wall shear stress, $\rho$ is the blood density, $r_i$ is the vessel internal radius, and $f$ is a linear or non-linear function of $A$.

If instead the coronary circulation is modelled through a 0D lumped parameter approach [2, 13, 14], the hemodynamics is modeled through a suitable combination of electrical counterparts: resistances ($R$) accounting for the viscous-dissipative effects, compliances/elastances ($C/E$) accounting for distensibility/elasticity effects, and inertances ($L$) accounting for inertial effects, giving rise to the most general 3-elements (or RLC) Windkessel model. Through the electric analogue, the only independent variable is time ($t$) and each vascular compartment is described by three time-dependent hemodynamic variables: pressure ($P$), flow rate ($Q$), volume ($V$). With reference to the RLC circuit reported in Fig. 1, each *i-th* compartment is characterized by three ordinary differential equations: an equation for the conservation of mass (expressed in terms of volume variation), a momentum equation (accounting for the flow variation), and a linear state equation between pressure and volume. Namely:

(2)
$$\begin{cases} \dfrac{dV_i}{dt} = Q_{i-1} - Q_i \\ \dfrac{dQ_i}{dt} = \dfrac{P_i - P_{i+1} - R\,Q_i}{L} \\ \dfrac{dP_i}{dt} = \dfrac{1}{C}\dfrac{dV_i}{dt} \end{cases}$$

where the subscripts *i+1* and *i-1* refer to the compartments downstream and upstream to the current *i-th*, respectively.

Equations (1) e (2) represent the general and common form for the 1D and 0D hemodynamic modeling, respectively. For the sake of simplicity and space, we neglect details of the different modeling choices adopted in literature, which however are formulations obtained starting from the above set of equations (1) and (2);

- Upstream coupling with systemic circulation is usually handled by 1D or 0D models [2, 13, 14] as described in Eq. (1) and (2). Typical flow rate and pressure signals at the aortic root are sometimes directly imposed as inlet boundary conditions for the coronary circulation;
- Downstream coupling with distal coronary circulation is in most of the cases dealt with 3-elements Windkessel models [2, 13, 14] as described in Eq. (2) and sometimes with a single impedance element;
- Mechanical coupling between the myocardial tissue and coronary walls can be included to account for the vessel squeezing during the heart contraction phase [10, 12, 15, 16];
- Different mechanisms of myocardial blood flow regulation, such as myogenic, flow, and metabolic control, can be added to the coronary modeling [12, 17, 18].

Both mechanical heart contraction coupling and autoregulation mechanisms are modeled through algebraic or ordinary differential equations, which are coupled to the arterial coronary model.

## 3. Results

We here report a chronological overview of 0D-1D coronary blood flow modeling appeared in the last two decades, by mainly focusing on the methodological solutions adopted, results, and aim of each study. Then, recent applications of 1D coronary modeling in presence of cardiac arrhythmias are described more in detail.

*3.1 Overview of 0D-1D coronary blood flow modeling*

We start with the paper by Smith et al. (2002) [19], which is a seminal and numerically-oriented work in this area. The authors implemented a very detailed and artificially-built coronary network through a classical 1D model, with parabolic velocity profile, elastic walls and no time dependent resistances to account squeezing due to heart contraction. Distal coronary circulation was modeled through appropriate 0D models. The aim of the paper was to describe a mathematical model and discuss its numerical properties, through idealized configurations, such as propagation along the network of an impulsive pressure input and wash-out of the network. The results showed physiologically realistic flow rates, washout curves, and pressure distributions.

Huo and Kassab (2007) [20] adopted a classical 1D model for the main coronary arteries (RCA, LAD and Cx) and corresponding primary branches, with elastic arterial walls. Attention was paid to dissipation: an additive nonlinear dissipative term was included in the momentum equation, and energy dissipation was considered at bifurcations. The inlet pressure boundary condition was obtained from experimental measurements, while an assigned impedance was imposed as outlet boundary condition for the downstream distal coronary circulation. The aim of the authors was to show the reliability of the 0D-1D model against measured data. Nonlinear convective term and energy loss at bifurcations were found to play a negligible role in the larger epicardial vessels of an arrested heart. Important outcomes also came from the scaling of the flow waves along the main coronary arteries and primary branches.

Van Der Horst et al. (2013) [15] implemented a classical though quite detailed 1D model for the coronary arteries, adopting ideal Murray law [14] at bifurcations but with possible inclusion of patient-specific physiological and geometric clinical data, in addition to the generic healthy subject configuration. Approximated Womersley velocity profile (i.e., pulsatile flow with non-parabolic velocity profile and varying over a sinusoidal cycle [14]) was adopted and 3-elements Windkessel models mimicked the distal coronary region. Accurate description of the mechanics of heart tissue and coronary walls was carried out. The squeezing emerged naturally by the detailed model of the stress state in myocardium and arteries. The authors goal was to propose a model where the myocardial tissue mechanics is well described. Two diseases were modeled: coronary epicardial stenoses and left ventricular hypertrophy with an aortic valve stenosis. The simulations showed that the model adequately reproduces the coronary hemodynamics (in terms of pressure and flow rate) in both healthy and diseased cases.

In a first paper, Mynard et al. (2014) [21] proposed a classical, anatomically-based and very detailed 1D model for the left arterial coronary circulation (LAD and Cx), with wall viscoelastic correction and a 0D model for each intramyocardial microvascular bed downstream of each 1D coronary artery. The 0D model was divided into three transmural layers (subendocardium, midwall, and subepcardium), while autoregulation mechanisms related to the intramyocardial pressure and myocardial filling were as well accounted for. In a later paper of 2015 by Mynard et al. [22], the coronary modeling considered only the main coronary arteries and branches, was extended to the right coronary circulation and coronary veins, and was eventually included into a 0D-1D model of the entire adult circulation. The purpose was to present and *in vivo* validate a coronary model able to describe microvascular properties that differ regionally and transmurally, as well as wave propagation effects in the conduit arteries. The authors were able to quantify forward and backward wave intensity as well as the interaction between cardiac function/mechanics and wave dynamics. No greater prominence of wave propagation effects was found in adults compared with newborns.

Rivolo et al. (2016) [23] adopted a classical 1D model for the arterial coronary circulation, with complex and very detailed vessel network geometry extracted by real data. Arterial walls were modeled as linear elastic, while the myocardium (at distal position in the network) was modelled as porous medium. 1D model was run on these networks and wave patterns and reflection coefficients were investigated along the three main coronary arterial paths (LAD, Cx, RCA). The authors aimed at investigating the forward and backward wave pattern in coronaries and discussing the scaling law for wave transmission at bifurcations, with attention to the role of pulsatility. Results demonstrated that, under different pulse wave speed conditions considered, the absence of reflection for pressure/flow waves traveling from the coronary stem towards the microcirculation is a salient feature of the coronary vasculature.

Based on [21], Guala et al. (2017) [24] implemented a classical 1D model for the main arteries and branches of the left arterial coronary circulation (LAD and Cx), with explicit non-linear viscoelastic wall properties and 0D models of intramyocardial vascular beds downstream of the 1D circulation. The 0D-1D coronary is upstream coupled with the left heart and systemic arterial tree, which is modeled as the large coronary conduits (i.e., by 1D models with non-linear viscoelastic arterial wall). The aim was to study age-induced effects on the wave propagation and reflection, as well as on the ventricular function. The authors found that the subendocardial viability ratio decreases with age, the total coronary flow slightly reduces, and the left ventricular work increases, resulting in a possible oxygen supply-demand unbalance due to physiological aging.

The works by Ge et al. in 2018 [25, 26] are as well based on the models by Mynard et al. [21, 22], with viscoelastic arterial walls and specific model for energy dissipation in stenosis. Arterial coronary circulation, constituted by 87 large coronary arteries (LAD, Cx, RCA, and 53 penetrating arteries), was represented by classical 1D models coupled with 0D models of intramyocardial microcirculation beds. Vessels in each intramyocardial vascular subsystem were distributed in multiple myocardial layers and represented by a series of variable resistances, compliances, and inductances, to account for the effects of the time-varying and depth-dependent intramyocardial tissue pressure on vascular deformation and blood flow. A flow autoregulatory mechanism was introduced to simulate the regulation of myocardial flow in the presence of coronary artery stenosis. Aim of these papers was twofold: (i) to quantitatively evaluate the sensitivity of transmural flow distribution to various cardiovascular and hemodynamic factors; (ii) to investigate the coronary flow autoregulation in the presence of coronary artery stenosis. The results demonstrated the interactive effects of HR, intramyocardial tissue pressure, and coronary perfusion pressure on transmural myocardial flow. Moreover, the sensitivity of coronary flow reserve to microvascular dilatation function and HR were found to be dependent on the severity of coronary stenosis.

Duanmu et al. (2019) [27] adopted a classical 1D model for the three main coronary arteries (LAD, Cx, RCA) and branches, based on angiography data from one subject. Distal coronary circulation was implemented with structured trees (i.e., artificial 1D networks) describing vessels in the myocardium. Inlet pressure was assigned according to a time-varying elastance model. Coronary squeezing was simulated by a time-dependent feedback pressure that was added to the distal boundary conditions for each terminal large artery. Aim of the study was the wave intensity analysis to simulate blood flow in the developed coronary model and the effects of arterial wall stiffness in the Cx. Pathological cases, such as stiffer large arteries and coronary epicardial diseases, were also investigated. Wave intensity analysis was found to properly distinguish different pathological reactions of the cardiovascular system. In particular, larger feedback pressure in terminal vessels increased the backward wave and decreased the forward one.

The works by Fan et al. in 2020 [28] and 2021 [29] differ from the classical 1D model as each coronary vessel was treated by a three-element Windkessel model, with two nonlinear resistances and a compliance, accounting for the vessel radius and the transvascular pressure. The model for the left coronary perfusion (LAD and Cx) consists of 400 vessels with 195 bifurcations, and 79 terminal vessels. The coronary model was coupled with the systemic circulation and the left ventricle contraction-relaxation mechanism, by considering the interplay between coronary perfusion and the corresponding regional contractility. The most recent paper [29] also included three mechanisms of flow regulation: shear response (inducing relaxation of the vessel radius due to changes in the vascular wall shear stress), myogenic responses (i.e., the ability of vascular smooth muscle to constrict a vessel in response to an increase of the transvascular pressure) and flow regulation controlled by a vasodilator signal related to physiological necessity. Aim of the two works was to investigate the mechanical dyssynchrony, namely the contractile dyssynchrony between the septum and left ventricular free wall due to asynchronous activation of the heart. The outcomes revealed that regional intramyocardial pressures play a significant role in affecting regional coronary flows in mechanical dyssynchrony. In particular, mechanical dyssynchrony through asynchronous activation influences coronary flow reserve and coronary flow, thus reducing the global contractility and left ventricular function.

Papamanolis et al. (2021) [30] coupled a classical 1D arterial coronary model to a single-compartment porous (Darcy) model for the myocardium. The vascular network of aorta and large epicardial coronary vessels (LAD, Cx, RCA) was obtained from CT angiography, while synthetic vessels were built for the downstream distal coronary circulation. The coronary network model coupled with the Darcy model mimicking myocardium gave satisfactory results against in-vivo data, also in case of impaired myocardial blood flow. Aim of the work was to provide a multiscale patient-specific model enabling blood flow simulation from large coronary arteries to myocardial tissue. Numerical results compared well with *in vivo* data and, in case of severe obstructive disease, related coronary artery narrowing with impaired myocardial blood flow, demonstrating the model ability to predict myocardial regions with perfusion deficit.

We conclude this overview by briefly mentioning few recent works, where coronary blood flow was modeled through a fully lumped parameter (0D) approach, by means of electrical counterparts adequately tuned on the basis on measured data. The aim of these studies was diversified but in all cases quite specific: to obtain a personalized framework for the coronary perfusion [31, 32], to have a faster computational tool for planning coronary artery bypass surgery [33], and to focus on myocardial mechanical function coupled with coronary perfusion [16].

*3.2 Multiscale coronary modeling: application to cardiac arrhythmias*
Gamilov et al. (2020) [34] implemented a classical and quite detailed 1D model of coronary circulation for the three main arteries (LAD, Cx, RCA) and corresponding branches, with elastic artery walls. Upstream boundary condition was assigned through a flow rate time series at the aorta root, while distal outflow boundary conditions assumed that a terminal artery is connected to the venous pressure reservoir with an assigned constant pressure. Different cardiac feedbacks on coronary flow were included: (i) dependency of the length of systole on the HR; (ii) linear dependency of the SV on the HR; and (iii) time-dependence of peripheral resistances on heart cycle. The authors aimed at studying coronary blood flow in the cases of interventricular dyssynchrony, tachycardia, bradycardia, long QT syndrome and premature ventricular contraction. The assigned discharge at aorta root was always modelled as a sinusoidal function. The investigated cases were obtained by deterministically changing the frequency and the amplitude of such function, while no stochastic term was introduced. The authors found that coronary blood flow (CBF), which is the net blood flow through coronary arteries (LCA and RCA) per beat, was significantly affected by arrhythmias. In particular, with respect to the baseline pacing case (which is here reported in Fig. 2a for the overall CBF, accounting for LCA+RCA), for long QT syndrome CBF decreased at rest (60 bpm) by 26% in LCA and 22% in RCA (see Fig. 2b). During bigeminy, trigeminy, and quadrigeminy, overall CBF decreased by 28%, 19%, and 14% with respect to the baseline pacing at rest (60 bpm), respectively (see Fig. 2c).

Scarsoglio et al. (2019) [35] proposed a classical 1D coronary model based on [22]. The left arterial coronary circulation (LAD, Cx, and their main branches) was simulated through 1D models with explicit non-linear viscoelastic wall properties, while 0D models of intramyocardial vascular beds were imposed as downstream distal boundary conditions of the 1D circulation. A combined 0D-1D model of the left heart and systemic arterial tree was imposed as upstream boundary condition of the coronary model, with systemic arteries modeled as the large coronary conduits (i.e., by 1D models with non-linear viscoelastic arterial wall). AF was simulated at different ventricular rates (50, 70, 90, 110, 130 bpm) by artificially-built RR stochastic extraction which mimics the *in vivo* beating features (i.e., faster, more variable and temporally uncorrelated RR). Aim of the work was to explore whether and to what extent ventricular rate during AF affects the coronary perfusion. The authors focused on AF-induced changes related to increasing HR on the waveform, amplitude and perfusion of the coronary blood flow at the LAD level. Waveform and amplitude of the LAD flow rate were strongly modified by the HR increase and beat-to-beat variability. Higher HR during AF exerted an overall coronary blood flow impairment and imbalance of the myocardial oxygen supply-demand ratio, as shown in Fig. 3a, where the CBF and rate pressure product (RPP) are displayed as function of the HR. The combined increase of HR and higher AF-induced hemodynamic variability led to a coronary perfusion impairment exceeding 90–110 bpm in AF. It was also found that coronary perfusion pressure (CPP) is no longer a reliable proxy of myocardial perfusion for HR higher than 90 bpm, see Fig. 3b.

Saglietto et al. (2021) [36] of the same research group as in [35], recently proposed a study with further modeling improvements aimed at deepening the comprehension of AF effects on coronary perfusion across the myocardial layers. The left and right arterial coronary circulation (LAD, Cx, RCA, and their main branches) was implemented similarly to [22], through 1D models with explicit non-linear viscoelastic wall properties. A 0D model, divided into three transmural layers (subendocardium, midwall, and subepcardium), was adopted as distal condition for each penetrating vasculature and microcirculatory district downstream of each 1D coronary artery. The coronary model was included into a 0D-1D model of the entire adult circulation, accounting for 1D modeling of the systemic arterial tree (which served as the upstream boundary condition for the coronary circulation), and a 0D lumped parameter representation of the venous return, contractile heart, and pulmonary circulation. The model also included a short-term baroreceptor mechanism accounting for the control of the systemic vasculature, chronotropic and inotropic effects. AF was simulated by stochastic beating extraction as in [35], with both atrial elastances constant to mimic the absence of atrial kick. Three mean HR were simulated (75, 100, 125 bpm) in both sinus rhythm (SR) and AF. An inter-layer and inter-frequency analyses were conducted downstream of the three main coronary arteries (LAD, Cx, RCA), by focusing on the ratio between mean beat-to-beat blood flow in AF compared to SR, $\overline{Q}_{AF}/\overline{Q}_{SR}$. As reported in Fig. 4, AF caused a direct reduction in microvascular coronary flow particularly at higher HR, with the most important decrease seen in the subendocardial layers perfused by LAD and Cx.

## 4. Discussion

The 0D-1D computational modeling of coronary circulation here reviewed highlights interesting findings and cues, both for future modeling improvements and clinical applications. If 1D coronary modeling is quite analogous in literature data (only arterial network details usually change), the variety of solutions adopted to handle upstream and downstream boundary conditions, as well as feedbacks from heart contraction and coronary flow regulation, shows that multiscale modeling is versatile for different clinical applications and represents an increasingly useful tool in translational medicine. On the other hand, the variety of modeling choices and the different purposes and applications simulated make a systematic comparison between literature studies not straightforward. Most of the computational results focused on the LAD, Cx and RCA, only [27, 34] analyzed more in details the left coronary artery (LCA) in comparison with its downstream branches, LAD and Cx. This is probably due to the fact that LCA is considered a conduit artery being much more close to the aortic root and less representative of the intralayer perfusion and the microvasculature.

In the specific case of cardiac arrhythmias, the analyzed studies agree in identifying a maximum of CBF (either computed on the LAD or overall on the LCA + RCA) around an average HR of 90-110 bpm. However, the reduced coronary perfusion beyond this HR threshold is not matched by a reduction of oxygen consumption, resulting in an unbalance of the coronary supply-demand ratio. QT syndrome appears to affect right and left coronary perfusion similarly, but to a greater extent the lower HRs. Instead, left coronary microvasculature is much more affected by AF than the right counterpart and this can be explained by the reduced driving pressure (aortic pressure) and the increased extravascular force (left ventricular end-diastolic pressure), both exacerbated by increasing HR. These latter mechanisms can also explain why the subendocardial layer is the most prone to AF-induced alterations. Both in AF and in quadrigeminy, trigeminy and bigeminy cases, coronary perfusion worsens for increasing HR.

Even though still small in number, computational studies already outline important clinical implications. First, the role of irregular RR beating in AF on the coronary perfusion is poorly understood. Several mechanisms - such as reduced aortic pressure related to short RR intervals, coronary vasoconstriction, and reduced coronary blood flow - have been proposed to explain how irregular ventricular rhythm can exert negative effects on the coronary hemodynamics. However, the precise mechanisms through which AF impacts the coronary circulation still remain to be explored [7]. Another clinical implication is related to the impact of ventricular rate on the coronary perfusion. Although lenient (resting HR < 110 bpm) and strict (resting HR < 80 bpm) rate control strategies were found not to differ in terms of mid-term cardiovascular outcomes [37], definitive data are still missing [38] and specific findings related to coronary flow impairment are overall scarce. For both rhythm and rate alterations, computational modeling can suggest useful evidences to clinical practice and better address further necessary *in vivo* investigation.

Given the frequent coexistence of AF and other cardiac diseases - such as coronary artery disease and heart failure [39, 40] - and the substantial absence of computational studies considering the concomitant presence of different cardiac pathologies, future modeling efforts should be devoted to investigate to what extent the interaction of AF with other cardiac diseases affects the coronary circulation.

## 5. Limitations

3D computational approaches were excluded from this review, as the focus was on 0D-1D models only. While the first ones are much more detailed but computationally expensive, being not feasible to describe the entire circulation but only the local hemodynamics, the latter represent a good settlement between the computational cost, the level of the hemodynamic details, and the reliability of the outcomes for the whole cardiovascular system. 0D-1D models are able to capture wave transmission and reflection phenomena and are well suited to study the effects of cardiac arrhythmias on the coronary circulation. In addition, the reported 0D-1D studies offer a diversified scenario of the modeling approaches adopted. If 1D coronary modeling is quite analogous in literature data (only arterial network details usually change), other physiological mechanisms, such as coronary flow regulation, viscolelastic arterial walls, heart contraction feedbacks and proper distal boundary conditions, are accounted for or not depending on the aim pursued. Thus, to date, an overall concordant framework of optimal modeling solutions cannot be easily identified. Last, the focus on cardiac arrhythmias led to exclude investigations mostly aimed at different goals, such as surgery procedures, coronary flow regulation, and other cardiovascular diseases.

## 6. Conclusions

The literature reviewed in the current work highlights a lively research activity related to 0D-1D models focused on coronary hemodynamics. In particular, cardiac arrhythmias by their own have been shown to promote relevant alterations of the coronary hemodynamics. The computational approaches here discussed are not merely able to describe the direct hemodynamic effects of cardiac arrhythmias, but also provide unique hints regarding the possible mechanisms behind these phenomena. The current small number of coronary models applied to cardiac arrhythmias stresses the need for future studies focused on this topic. The present review can boost future research devoted to the hemodynamic effects exerted by AF and other arrhythmias on the coronary circulation, exploiting the great potential of multiscale mathematical modeling in cardiovascular fluid dynamics.

## Author contributions

SS and LR conceived and designed the review study, analyzed the literature data, wrote, revised and approved the final version of the manuscript.

## Ethics approval and consent to participate

Not applicable.

## Acknowledgments

We would like to express our gratitude to all the peer reviewers for their opinions and suggestions.

## Funding

This research received no external funding.

## Conflict of interest

The authors declare no conflict of interest.

| | | | |
|---|---|---|---|
| **A** | **Cross-sectional area** | **LAD** | **Left anterior descending artery** |
| **AF** | **Atrial fibrillation** | **LCA** | **Left coronary artery** |
| **C** | **Compliance** | **MID** | **Midwall** |
| **CAD** | **Coronary artery disease** | **P** | **Blood pressure** |
| **CBF** | **Coronary blood flow** | **Q** | **Blood flow rate** |
| **CPP** | **Coronary perfusion pressure** | **R** | **Resistance** |
| **Cx** | **Left circumflex artery** | **RCA** | **Right coronary artery** |
| **E** | **Elastance** | **RPP** | **Rate pressure product** |
| **ENDO** | **Subendocardium** | **RR** | **Cardiac beating period** |
| **EPI** | **Subepicardium** | **SR** | **Sinus rhythm** |
| **HR** | **Heart rate** | **SV** | **Stroke volume** |
| **L** | **Inertance** | **V** | **Blood volume** |

**Table 1. Legend of the abbreviations and symbols.**

**Figure legends**

Figure 1. Conceptual scheme for state of the art 0D-1D coronary blood flow modeling. Symbols are defined in Table 1.

Figure 2. Results by Gamilov et al. (2020) [34]: (a) Overall CBF (LCA+RCA) [ml/s] as function of HR for baseline pacing; (b) Ratio of the CBF for LCA and RCA with respect to baseline pacing for the QT syndrome case; (c) Ratio of the overall CBF (LCA+RCA) with respect to baseline pacing in cases of quadrigeminy, trigeminy, bigeminy. Symbols are defined in Table 1.

Figure 3. Results by Scarsoglio et al. (2019) [35]: (a) Mean values (μ, solid curves) and standard deviation values (in terms of μ±σ, shaded areas) of CBF (left y-axis), RPP (right y-axis) as function of HR; (b) coefficient of determination, $R^2$, for the relation CBF(CPP) as function of HR. Symbols are defined in Table 1.

Figure 4. Results by Saglietto et al. (2021) [36]: $\overline{Q}_{AF}/\overline{Q}_{SR}$ at each HR across the myocardial layers (EPI, MID, ENDO), in three downstream microvascular districts: (a) LAD, (b) Cx, (c) RCA. Rate-specific regression lines (dashed lines) are also reported. Symbols are defined in Table 1.

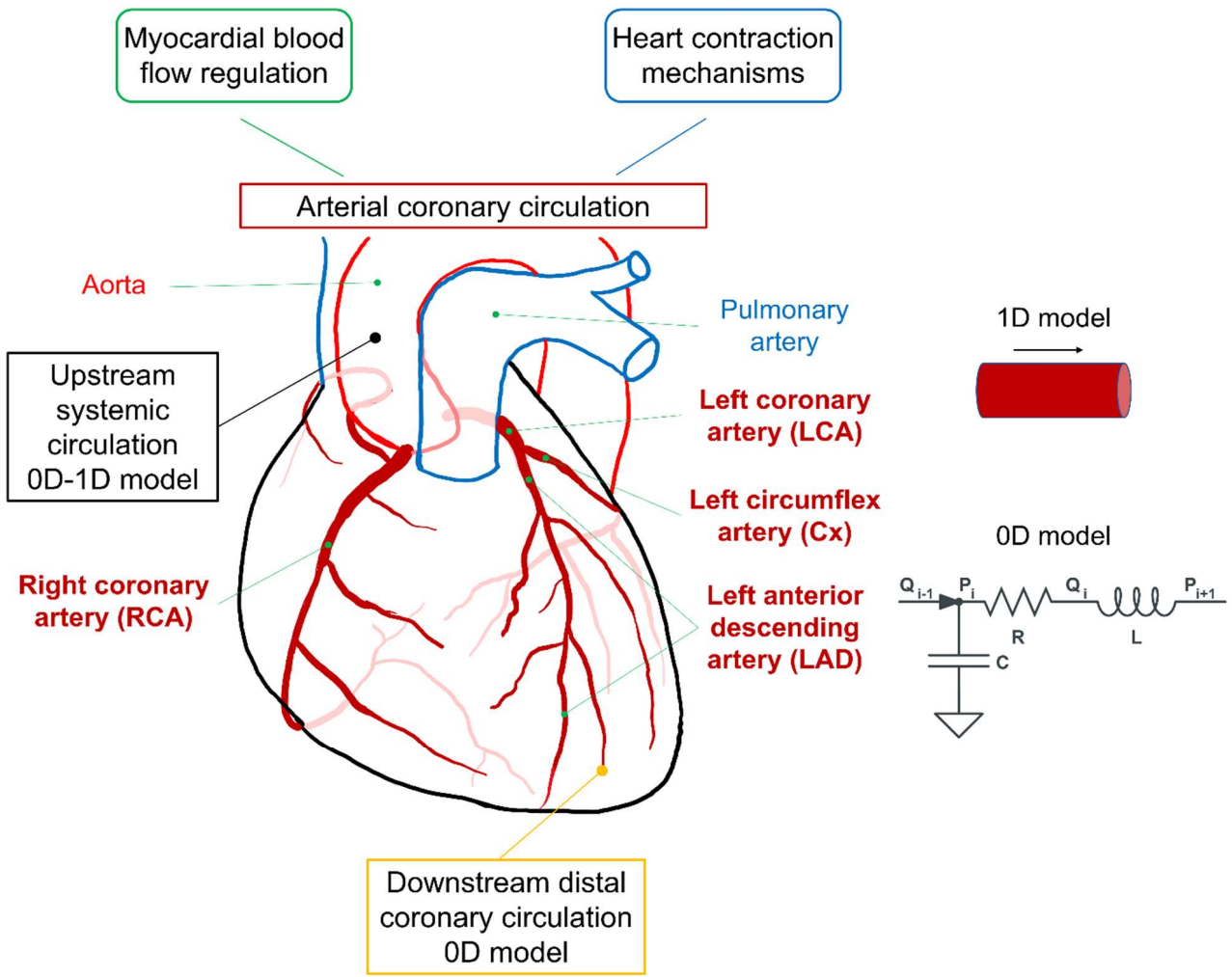

**Figure 1.**

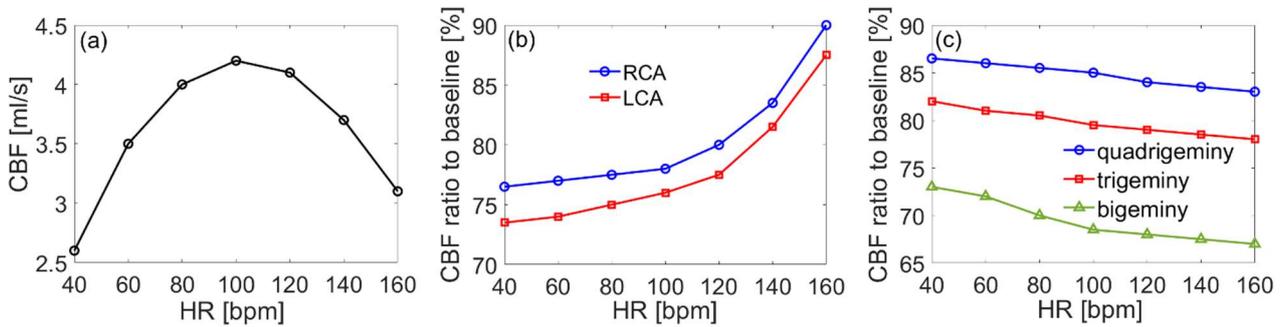

**Figure 2.**

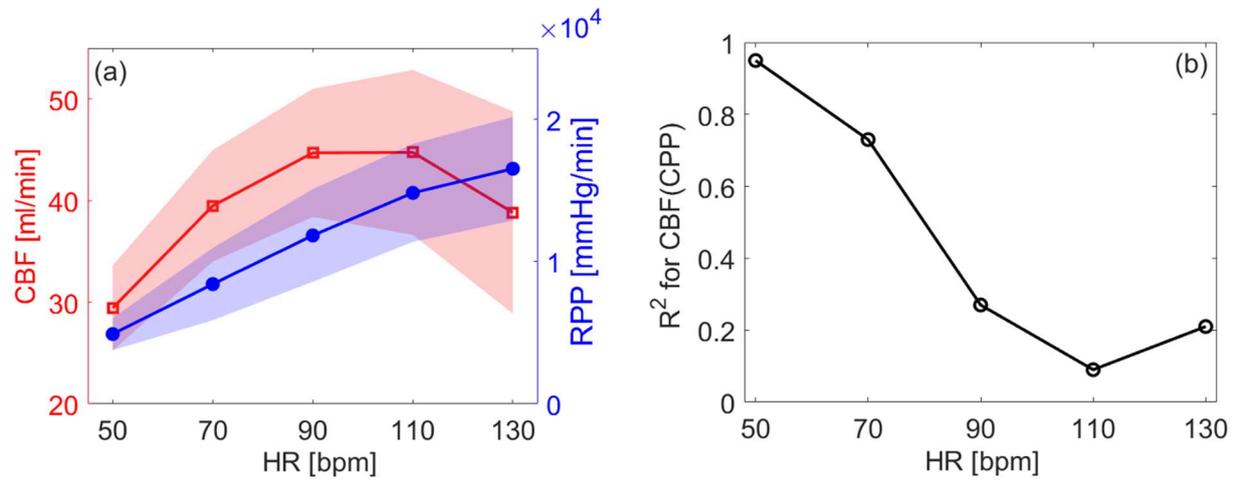

**Figure 3.**

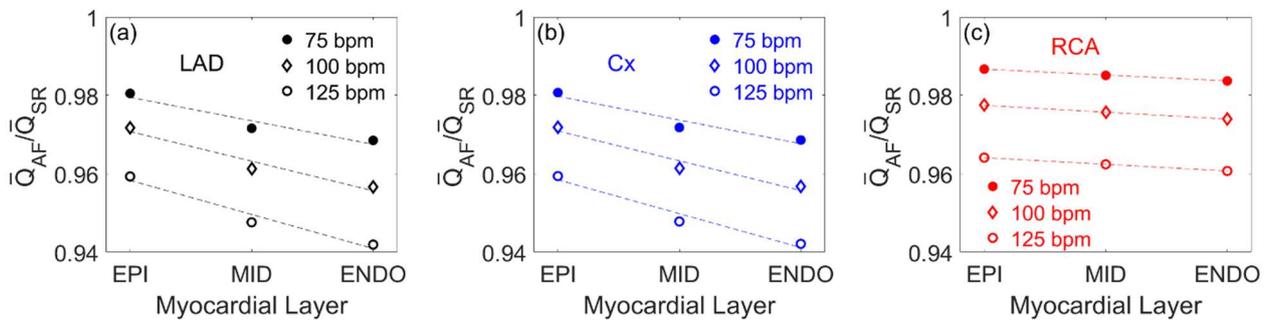

**Figure 4.**